\documentclass[prl,twocolumn,showpacs,superscriptaddress]{revtex4}
\usepackage[dvips]{graphicx}
\usepackage{hyperref}
\usepackage{bm}

\begin{document}

\def\jpb{J. Phys. B: At. Mol. Opt. Phys.~}
\def\pra{Phys. Rev. A~}
\def\prb{Phys. Rev. B~}
\def\prl{Phys. Rev. Lett.~}
\def\jmo{J. Mod. Opt.~}
\def\jetp{Sov. Phys. JETP~}
\def\etal{{\em et al.}}

\def\reff#1{(\ref{#1})}

\def\Up{U_p}
\def\omegap{\omega_p}
\def\omegaMie{\omega_M}
\def\diff{\mathrm{d}}
\def\imagi{\mathrm{i}}

\def\halb{\frac{1}{2}}

\def\beq{\begin{equation}}
\def\eeq{\end{equation}}

\def\energy{{\cal E}}

\def\Ehat{\hat{E}}
\def\Ahat{\hat{A}}

\def\ket#1{\vert #1\rangle}
\def\bra#1{\langle#1\vert}
\def\braket#1#2{\langle #1 \vert #2 \rangle}

\def\vekt#1{\bm{#1}}
\def\vect#1{\vekt{#1}}
\def\vektr{\vekt{r}}
\def\vektp{\vekt{p}}
\def\vektE{\vekt{E}}
\def\vektEhat{\hat{\vekt{E}}}
\def\vektB{\vekt{B}}
\def\vektv{\vekt{v}}
\def\vektk{\vekt{k}}
\def\vektkhat{\hat{\vekt{k}}}

\def\makered#1{{\color{red} #1}}

\def\Im{\,\mathrm{Im}\,}

\def\varphic{\varphi_{\mathrm{c}}}

\def\vxc{v_\mathrm{xc}}
\def\vextop{\hat{v}_\mathrm{ext}}
\def\VC{V_\mathrm{c}}
\def\VHX{V_\mathrm{Hx}}
\def\VHXC{V_\mathrm{Hxc}}
\def\wop{\hat{w}}
\def\Gammaevenodd{\Gamma^\mathrm{even,odd}}
\def\Gammaeven{\Gamma^\mathrm{even}}
\def\Gammaodd{\Gamma^\mathrm{odd}}

\def\Hop{\hat{H}}
\def\HopKS{\hat{H}_\mathrm{KS}}
\def\HKS{H_\mathrm{KS}}
\def\Top{\hat{T}}
\def\TopKS{\hat{T}_\mathrm{KS}}
\def\VopKS{\hat{V}_\mathrm{KS}}
\def\VKS{{V}_\mathrm{KS}}
\def\vKS{{v}_\mathrm{KS}}
\def\Ttildeop{\hat{\tilde{T}}}
\def\Ttilde{{\tilde{T}}}
\def\Vextop{\hat{V}_{\mathrm{ext}}}
\def\Vext{V_{\mathrm{ext}}}
\def\Vopee{\hat{V}_{{ee}}}
\def\psiopdag{\hat{\psi}^{\dagger}}
\def\psiop{\hat{\psi}}
\def\vext{v_{\mathrm{ext}}}
\def\Vee{V_{ee}}
\def\vee{v_{ee}}
\def\nop{\hat{n}}
\def\Uop{\hat{U}}
\def\Wop{\hat{W}}
\def\bop{\hat{b}}
\def\bopdag{\hat{b}^{\dagger}}
\def\qop{\hat{q}}
\def\jop{\hat{j\,}}
\def\vHxc{v_{\mathrm{Hxc}}}
\def\vHx{v_{\mathrm{Hx}}}
\def\vH{v_{\mathrm{H}}}
\def\vc{v_{\mathrm{c}}}
\def\xop{\hat{x}}

\def\Wcmcm{W/cm$^2$}

\def\varphiexact{\varphi_{\mathrm{exact}}}

\def\fmathbox#1{\fbox{$\displaystyle #1$}}

\title{Relativistic Attosecond Electron Bunches from Laser-Illuminated Droplets }

\author{T.V.~Liseykina}
\affiliation{Max-Planck-Institut f\"ur Kernphysik, Postfach 103980, 69029 Heidelberg, Germany}
\affiliation{Institute for Computational Technologies, SD-RAS, Novosibirsk, Russia}
\author{S.~Pirner}
\affiliation{Max-Planck-Institut f\"ur Kernphysik, Postfach 103980, 69029 Heidelberg, Germany}
\author{D.~Bauer}
\affiliation{Max-Planck-Institut f\"ur Kernphysik, Postfach 103980, 69029 Heidelberg, Germany}

\date{\today}

\begin{abstract} The generation of relativistic attosecond electron bunches is observed in three-dimensional, relativistic particle-in-cell simulations of the interaction of intense laser light with droplets. The electron bunches are emitted under certain angles which depend on the ratios of droplet radius to wavelength and plasma frequency to laser frequency. The mechanism behind the multi-MeV attosecond electron bunch generation is investigated using Mie theory. It is shown that the angular distribution and the high electron energies are due to a parameter-sensitive, time-dependent local field enhancement at the droplet surface.    
\end{abstract}
\maketitle

The understanding of the laser energy conversion into fast electrons and ions is of utmost importance for the design of efficient ``table-top'' particle accelerators \cite{gibbonbook,mourou,salamin}. The laser plasma-based acceleration schemes discussed so far in the pertinent literature may be divided into two groups according to whether the plasma is underdense, i.e., the plasma frequency $\omegap$ is smaller than the laser frequency $\omega$, or vice versa. Wake-field accelerators and the so-called ``bubble-regime'' (see, e.g.,  \cite{leemans,dreambeams,pukhov}) fall into the former category while the interaction of intense laser pulses with solid surfaces or thin foils (see e.g., \cite{brandl,kar,fuchs,green,macchi}) belongs to the latter, overdense regime.
   
Of particular interest are finite-size targets where fast particles cannot escape into the field-free bulk material but yet the density of the accelerated particles may be sizable.  In recent years the nonrelativistic interaction of intense laser light with small, subwavelength-size clusters has been thoroughly investigated \cite{saalmannandfennel}. In particular, the efficient absorption of laser energy, leading to high charge states and thus to intense line emission as well as to  energetic electrons and ions was studied. In such small clusters of radii $R<\delta\ll\lambda$, where $\lambda$ is the laser wavelength and $\delta$ is the skin depth of the cluster plasma, the effect of the cluster on the propagation of the laser pulse needs not be taken into account. Although the plasma, which is created via field ionization on a sub-cycle time scale, is overdense, screening of the cluster interior only occurs due to polarization but not due to a skin effect. Technically speaking, the dipole approximation can be applied to the nonrelativistically intense laser field, $\vektE(\vektr,t)\simeq\vektE(t)$. As a consequence the electron dynamics mainly occurs in the laser polarization direction while the $\vektv\times\vektB$-force in laser propagation direction and the influence of the scattered electromagnetic field on the particle dynamics can be safely neglected. 

The other extreme of intense laser-matter interaction is constituted by targets of sizes much larger than a wavelength, e.g., a laser beam impinging on a solid surface. The absorption mechanisms in this case have also been extensively investigated \cite{lasersolid}. It is well-known, for instance, that for perpendicular incidence electrons are accelerated in laser propagation direction (i.e., into the bulk material). 

In the present Letter we will focus on electron acceleration in the regime where $R$ and $\lambda$ are of the same order of magnitude (i.e., rather droplets than clusters) and the laser intensity is relativistic, i.e., the ponderomotive energy $\Up$ exceeds the electron rest energy $mc^2$. 
By changing the droplet radius from $R\ll\lambda$ (small-cluster limit) to $R\gg\lambda$ (solid-surface limit) the emission angle of electrons is expected to decrease from $\theta=\pi/2$ to $0$ with respect to the propagation direction $\vektkhat$. Moreover, the emission is expected to occur in the ($\vektEhat,\vektkhat$)-plane for a linearly polarized laser field of amplitude $\vektEhat$.

\begin{figure}
\includegraphics[width=0.35\textwidth]{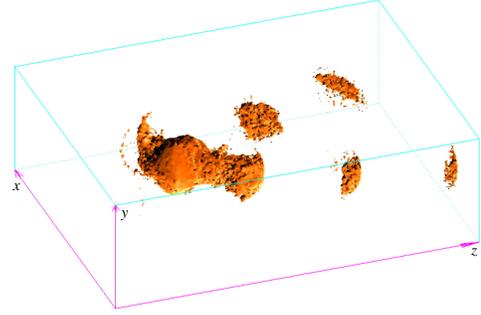} 
\caption{(Color online) Electron isocontour surfaces (1\% of $n_{e0}$) of a $R=\lambda/4$ He-droplet of density $n_{e0}=22\,n_c$ with $n_c=1.8\times 10^{21}$\,cm$^{-3}$ in a $16$-cycle $\sin^2$-laser pulse ($\lambda=800$\,nm) of intensity $2\times 10^{18}$\,\Wcmcm\ ($a=1$)  at $t=10$\,cycles. The laser pulse propagates in $z$-direction, the electric field is directed along $x$.   \label{3d_isosurface}}
\end{figure}

For a sphere in a plane electromagnetic wave all angles of incidence occur simultaneously. It is known from laser-plasma interaction studies that, depending on the plasma scale length and the laser intensity, there exists an optimal angle of incidence for the absorption of laser energy \cite{ruhlandmulser}: the steeper the plasma gradient the more this optimal angle is pushed towards $\pi/2$, i.e.\ grazing incidence. For sufficiently large scale lengths resonance absorption (see, e.g., \cite{gibbonbook}) can occur most efficiently under a certain optimal angle. Indeed, this effect, resulting in the electron emission {\em opposite} to the incoming laser pulse  has been observed in experiments with droplets and two-dimensional model simulations \cite{pengzheng}.  

For $R>\delta$ the self-consistent electromagnetic field needs to be calculated. Numerically we do this by means of three-dimensional, electromagnetic, relativistic particle-in-cell (PIC) simulations. In such simulations the particle dynamics, the deformation of the target, and the corresponding modification of the field is automatically taken into account. Analytically, the self-consistent electromagnetic field around and inside a laser-illuminated droplet of given dielectric permittivity $\varepsilon$ can be calculated using the corresponding solutions to Maxwell's equations put forward by Clebsch, Lorenz, Mie, Debye, and others. The electromagnetic scattering by dielectric or metallic spheres is commonly called ``Mie theory'' \cite{bornwolf,hulst,bohren}. Of course, the self-consistent particle {\em dynamics} far from equilibrium in very intense laser fields cannot be captured by a dielectric constant and thus is not included in Mie theory while it is included in the PIC simulations. However, we will show that Mie theory is nevertheless capable of explaining the angular distributions observed in the PIC simulations and the high electron energies exceeding the expected few times the ponderomotive energy. 

We start by presenting typical results from PIC simulations of the interaction of an intense plane wave laser pulse with a pre-ionized He droplet. 
Figure~\ref{3d_isosurface} shows a snapshot of the electron isocontour surfaces corresponding to 1\% of the initial electron droplet density. The dimensionless vector potential amplitude $a=\vert e\Ahat/mc\vert = \vert e\Ehat/m\omega c\vert$ of the  $16$-cycle $\sin^2$-laser pulse ($\lambda=800$\,nm) was $a=1$, corresponding to  a laser intensity $I\simeq 2\times 10^{18}$\,\Wcmcm.  Electron bunches emitted each half cycle under plus/minus a certain angle $\theta$ (with respect to the propagation axis $z$) in forward direction are clearly visible. The bunches are mainly confined to the plane of incidence (i.e., the $(x,z)$-plane).

\begin{figure}
\includegraphics[width=0.4\textwidth]{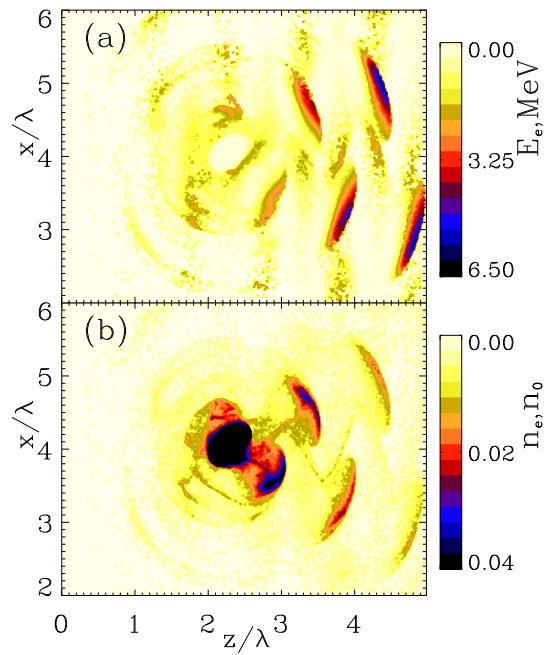} 
\caption{(Color online) Kinetic electron energy (a) and density (b) contour plots (in the $y=0$-plane)  of the He-droplet of Fig.~\ref{3d_isosurface}  in a $16$-cycle $\sin^2$-laser pulse ($\lambda=800$\,nm) of intensity $8\times 10^{18}$\,\Wcmcm\ ($a=2$)  at $t=8$\,cycles. \label{Energy_Density} }
\end{figure}

The electron energy and density for the same He-droplet in a four-times more intense laser pulse is presented in Fig.~\ref{Energy_Density}. Figure~\ref{Energy_Density}a shows that the bunches consist of electrons with energies up to $\simeq 6$\,MeV. The ponderomotive energy $\Up=mc^2(\sqrt{1+a^2/2}-1)$ is only $\simeq 0.37$\,MeV. As the spatial width of the bunches is much smaller than a wavelength, the temporal bunch structure is already in the attosecond domain. Similar bunches have been observed in PIC-simulations of the interaction of ultraintense laser pulses with solid surfaces under grazing incidence or with plasma channels \cite{naumova}. Such bunches have potential applications in attosecond electron diffraction experiments or the generation of coherent short-wavelength radiation via Thomson scattering. 

In the following we will explain the multi-MeV electron energies and their emission angles using Mie theory. The latter gives the analytical electromagnetic field configuration around and inside the (unperturbed) droplet in terms of series expansions over Legendre and Bessel functions \cite{bornwolf}. In the case of interest to us the droplet is conducting ($\varepsilon=1-\omegap^2/\omega^2$), the skin depth $\delta\simeq c/\omegap\ll R$, and the electric field on the droplet surface is perpendicular to it and quickly decays inside the droplet. Electrons at the droplet surface can be pulled out of the droplet or pushed inside, depending on the time they appear at the surface \cite{brandl}. 
\begin{figure}
\includegraphics[width=0.36\textwidth]{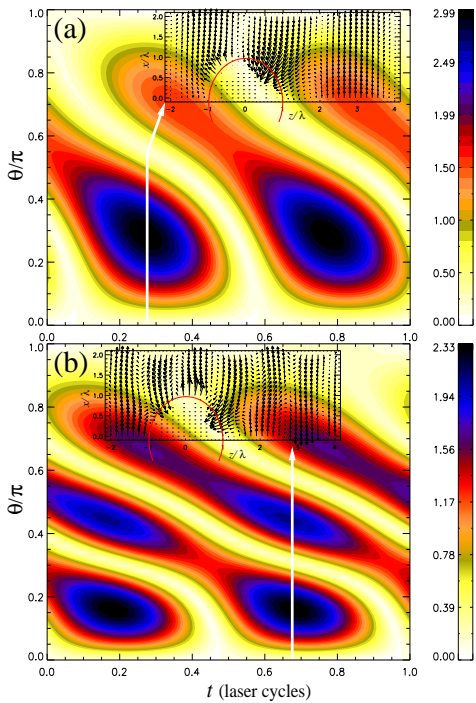} 
\caption{(Color) Absolute value of the radial electric field for a $22$ times overdense droplet at the droplet surface vs angle $\theta$ and time, as predicted by Mie theory for $R=\lambda/4$ (a) and $R=\lambda/2$ (b) in the $y=0$-plane. The color indicates the electric field in units of the incident field strength $\Ehat$. The insets show the electric vector field  in the $y=0$-plane at the ``optimal'' times (white arrows) when the electric field at the droplet surface is highest and pointing inwards, i.e.,  $t=0.275$~cyc.\ (a) and  $t=0.675$~cyc.\ (b), respectively. \label{mieI} }
\end{figure}
Figure~\ref{mieI} shows the absolute value of the radial electric field for an $\omegap^2/\omega^2=22$ times  overdense droplet at the droplet surface vs angle $\theta$ and time for $R=\lambda/4$ and $R=\lambda/2$ according Mie theory. Note that in our Mie-calculations the incident plane wave field is of the form $E_{\mathrm{inc},x}(z,t)=\Ehat\cos(kz-\omega t)$, i.e., has a constant amplitude. Figure~\ref{mieI} shows that Mie theory predicts certain times and angles at which the electric field at the surface is largest. Moreover, these electric field maxima may exceed the field amplitude $\Ehat$ (by a factor $3$ in Fig.~\ref{mieI}a and $2.3$ in  Fig.~\ref{mieI}b). For the bigger droplet the optimal angle is smaller, which means that the maximum electric field occurs more in propagation direction $\theta=0$. Secondary local maxima are visible at $\theta\simeq 0.45\pi$.

The two  insets in Fig.~\ref{mieI} show the electric vector field  in the $y=0$-plane according to Mie theory at the ``optimal'' times when the electric field at the surface is highest and points inwards (i.e., it pulls electrons outwards). In Fig.~\ref{mieI}a, for instance, $t\simeq 0.275$~cyc.\ and the largest field occurs under the angle $\theta=0.27\pi$. It is clear that under the same angle but $x<0$ the electric field has the same absolute value but points outward. As a consequence, electrons entering from inside the droplet into such a field configuration under this optimal angle will be efficiently accelerated outwards (inwards for $x<0$). Half a laser cycle later the situation reverses and electrons at $x<0$ will be accelerated outwards (inwards for $x>0$). This explains the observation of alternating electron acceleration into two directions each laser cycle.

\begin{figure}
\includegraphics[width=0.4\textwidth]{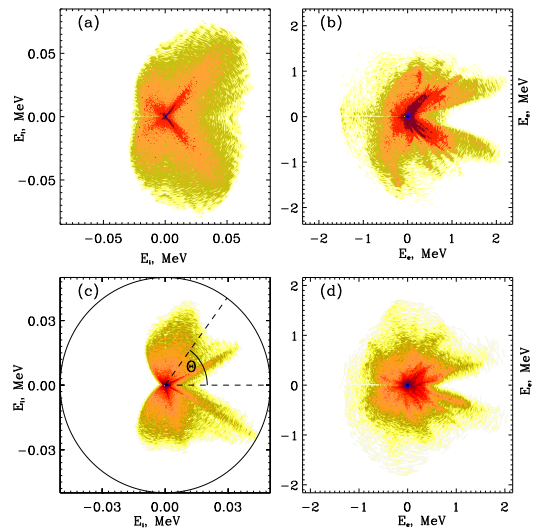} 
\caption{(Color online) Angle-resolved ion and electron kinetic energy spectra for the $R=\lambda/4$ droplet [panel (a) for ions and (b) for electrons] and the  $R=\lambda/2$ droplet [panel (c) for ions and (d) for electrons] after $t=9$~cycles in a $16$-cycle $\sin^2$-laser pulse ($\lambda=800$\,nm) of intensity $2\times 10^{18}$\,\Wcmcm\ ($a=1$). The emission angle $\theta$ [indicated in panel (c)] was determined as $\theta=\arctan(p_x/p_z)$ with $\vektp$ the momentum. The kinetic energy for electrons and ions reads $E_{e,i}=[m_{e,i}^2c^4+\vektp^2 c^2]^{1/2}-m_{e,i}c^2$. The color-coding is proportional to the logarithm of the particle number. \label{mieIII}}
\end{figure}

Figure~\ref{mieIII} shows angle-resolved ion and electron energy spectra obtained from the PIC simulations. The spectra were taken at time $t=9$~cycles when the fast electrons are still inside the simulation box. The ions are hardly set into motion at such an early time. However, the angular distribution of the ions in Figs.~\ref{mieIII}a and  \ref{mieIII}c have imprinted on them already the anisotropy due to the field distribution at the droplet surface. In fact, the emission angles are more easily inferred from the ion distributions than from the electron distributions in Figs.~\ref{mieIII}b and  \ref{mieIII}d because the electrons change their direction as they move away from the droplet. This is the reason why in Figs.~\ref{mieIII}b and  \ref{mieIII}d the energetic electrons are aligned stronger in forward direction than the slow electrons.

The emission angles inferred from the PIC results in Fig.~\ref{mieIII} are $\theta=0.27\pi$ for the $R=\lambda/4$-droplet and $\theta=0.14\pi$ for the  $R=\lambda/2$-droplet with a secondary maximum at $0.4\pi$. These results are in excellent agreement with the Mie results presented in Fig.~\ref{mieI}. We have performed a systematic study, comparing the emission angles predicted by Mie theory with those from the PIC calculations ($a=2$) for droplet radii between $\lambda/8$ and $\lambda/2$, and in all cases found very good agreement.

The unexpectedly high electron energies can be explained by the field enhancements at the droplet surface. As shown in Fig.~\ref{mieI}a the field amplitude at the surface of the $R=\lambda/4$-droplet is $3$ times the incident field amplitude. We have performed test-particle calculations where we placed electrons with zero initial velocity on the droplet surface at various phases with respect to the analytical, time-dependent  Mie field configuration. The maximum kinetic energies acquired by such test electrons are in good agreement with the maximum energies observed in the PIC simulations.

In the limit of small plasma droplets $kR=2\pi R/\lambda \ll 1$ the maximum radial field at the droplet surface predicted by Mie theory is
\beq E_r^{(\max)} = 3 \Ehat\ \frac{\omegap^2/\omega^2-1}{\omegap^2/\omega^2-3} . \label{Er}\eeq
For $\omegap^2\gg\omega^2$ one obtains $ E_r^{(\max)}/\Ehat=3$, i.e.\ a threefold field enhancement, as observed in Fig.~\ref{mieI}a (although $kR=\pi/2>1$ in this case). 
For, e.g., $\omegap=2\omega$, the predicted field enhancement according \reff{Er} is  $E_r^{(\max)}/\Ehat = 9$ so that one could argue that even more energetic electron bunches can be produced for lower-density droplets. However, only for $a\ll 1$ do the PIC calculations reproduce the predictions of Mie theory as far as the field enhancements at the droplet surface are concerned. This is because for relativistic laser intensities an only few-times overdense droplet quickly dissolves during the rising edge of the laser pulse.

\begin{figure}
\includegraphics[width=0.4\textwidth]{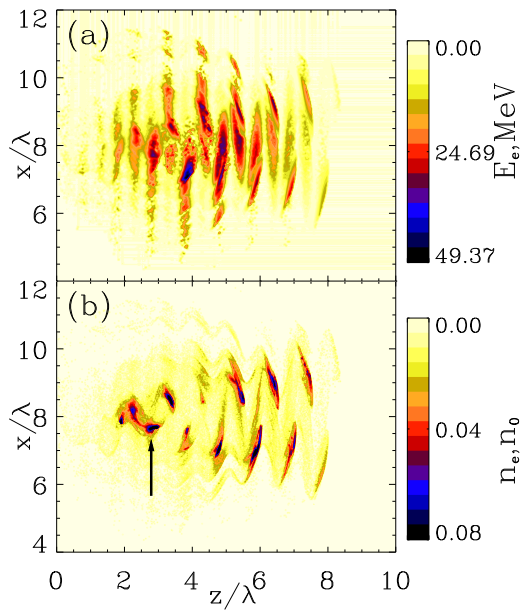} 
\caption{(Color online) Same as in Fig.~\ref{Energy_Density} but for $a=20$ ($8\times 10^{20}$\,\Wcmcm). The arrow in panel (b) indicates an electron bunch which, after the pulse, acquired an energy of $\simeq 130$\,MeV. \label{Energy_Density_strong} }
\end{figure}

Before concluding  we show the simulation result for the droplet of  Fig.~\ref{Energy_Density} in an ultra-intense laser field of intensity $8\times 10^{20}$\,\Wcmcm. At the plotting time ($t=8$\,cycles, i.e., at the maximum of the pulse) the highest electron energy observed is $\simeq 50$\,MeV. However, if one follows the electron bunch indicated by an arrow in  Fig.~\ref{Energy_Density_strong}b in time one finds a final energy of $\simeq 130$\,MeV, i.e., $20$ times the ponderomotive energy. From  Fig.~\ref{Energy_Density_strong}b one infers that almost all electrons are removed from the droplet. Nevertheless the angle under which the electron bunches are emitted is well-described by Mie-theory. However, the energetic electrons are not only located inside the electron bunches anymore at such high intensities, as is clearly visible in   Fig.~\ref{Energy_Density_strong}a.

In summary, we showed that multi-MeV attosecond electron bunches are produced when intense laser fields interact with overdense droplets of diameters comparable to the laser wavelength. The attosecond electron bunches are emitted each half laser cycle under plus/minus a certain angle in the polarization plane. The preferred electron emission angles and the high kinetic energies arise due to local field enhancements at the droplet surface that can be calculated using Mie theory. Relativistic attosecond electron bunches may be used for the generation of short-wavelength radiation via scattering of a counter-propagating laser pulse, time-resolved structural imaging, or plasma diagnostics. 

\bigskip

T.V.L.\ acknowledges support from the RFBR grants 09-02-12201 and 09-02-01103. S.P.\ is supported by the Studienstiftung des deutschen Volkes.

\end{document}